\documentclass[aps,prd,showpacs,amsmath,floats,floatfix,twocolumn,nofootinbib]{revtex4}
\usepackage{graphicx,amssymb,amsmath,amsbsy,mathrsfs}
\usepackage{bm}

\usepackage[usenames]{color}

\newcommand{\beq}{\begin{equation}}
\newcommand{\eeq}{\end{equation}}
\newcommand{\beqa}{\begin{eqnarray}}
\newcommand{\eeqa}{\end{eqnarray}}

\newcommand{\hsp}{\hspace{2mm}}
\newcommand{\hmm}{\hspace{1mm}}
\newcommand{\ovl}{\overline}

\newcommand{\brak}[2]{\left[ #1 , #2 \right]}

\begin{document}
\title{Degeneracy measures for the algebraic classification of numerical spacetimes}

\author{Robert Owen}
\affiliation{Center for Radiophysics and Space Research, Cornell University, Ithaca, NY 14853}
\email{owen@astro.cornell.edu}
\date{April 20, 2010}

\begin{abstract}
We study the issue of algebraic classification of the Weyl 
curvature tensor, with a particular focus on numerical 
relativity simulations.  The spacetimes of interest in this context, binary 
black hole mergers, and the ringdowns that follow them, present subtleties 
in that they are generically, strictly speaking, Type I, but in many regions 
approximately, 
in some sense, Type D.  To provide meaning to any claims of ``approximate'' 
Petrov class, one must define a measure of degeneracy on the space of null 
rays at a point.  We will investigate such a measure, used recently to argue 
that certain 
binary black hole merger simulations ring down to the Kerr geometry, after 
hanging up for some time in Petrov Type II.  In particular, we argue that this 
hangup in Petrov Type II is an artefact of the particular measure being used, 
and that a geometrically better-motivated measure shows a black hole merger 
produced by our group settling directly to Petrov Type D.  
\end{abstract}


\maketitle

\section{Introduction:}

The marvelous improvements in the technology of numerical relativity in recent 
years present opportunities for revolutionizing our understanding of the 
classical gravitational field.  In the past, much of this understanding has 
come from studying solutions with extreme symmetry, and perturbations of such 
solutions.  However, 
with the help of numerical methods, truly generic simulations, particularly of 
multiple black hole systems, can now be carried out in full general relativity.

While this work is undertaken, one must keep in mind the fundamental 
nature of general relativity and its solutions.  In particular, the general 
covariance of the theory is not naturally reflected in the numerical context, 
where gauge fixing is fundamentally required in the form of coordinate 
and tetrad choices.  In practice, such gauge choices are tailored to numerical 
convenience (or necessity), rather than to physical relevance.  Such a simple 
task as checking that a black hole merger settles down to a Kerr geometry 
can be clouded by the arbitrariness of the simulation coordinates.  

One way of dealing with these ambiguities would be to apply coordinate 
transformations to numerical simulations {\em a posteriori} to represent 
these spacetimes in physically preferable coordinates, if they exist.  If 
one needs to map 
all quantities to an entirely new coordinate grid, then some accuracy would 
presumably be lost to the interpolation process, especially if changes of the 
time function require interpolation in time.  More important, however, is the 
difficulty of fixing physically preferred coordinate systems in strongly 
dynamical and nonsymmetric spacetimes at all.

Another, perhaps complimentary, approach is to focus physical analysis on 
partially (or if possible, totally) gauge invariant quantities.  For example, 
a major tool in the analysis (and construction) of exact solutions in general 
relativity is the algebraic classification system of 
Petrov and Pirani~\cite{Petrov2000Reprint, Pirani1957, ExactSolutionsBook}, in 
which the Weyl tensor at any given point in spacetime is classified according 
to the algebraic properties of its associated {\em eigenbivector problem}:
\beq
{C_{ab}}^{cd} X_{cd} = \Lambda X_{ab}.\label{e:petrovproblem}
\eeq
Another view of this classification system, with a more geometrical 
flavor, was expounded particularly by Bel~\cite{Bel1962} 
and Penrose~\cite{Penrose1960}.  In this approach one classifies the Weyl 
tensor in terms of the degeneracy of the so-called {\em principal null 
directions}, null vectors defined up to scale by the equation:
\beq
k^e k^f k_{[a} C_{b] e f [ c} k_{d]} = 0.
\eeq
One can show (most easily in spinor language) that this equation is always 
satisfied by exactly four null rays, counting multiplicities.  If 
all four of these null directions are distinct, the spacetime is said to be 
{\em algebraically general} or {\em Type I} at that point in spacetime.  If 
two of them coincide, the spacetime is said to be {\em Type II} there.  If 
three, {\em Type III}.  If all four principal null directions coincide, the 
spacetime is said to be {\em Type N}, or {\em null}, in analogy with the pure 
radiation fields 
of vacuum electrodynamics.  If the principal null directions coincide in two 
distinct pairs, then the spacetime is said to be {\em Type D}.  The Kerr and 
Schwarzschild geometries are famous examples of globally Type D spacetimes, 
so in some sense, one may hope to infer that a spacetime is ``settling down 
to an approximate Kerr geometry'' if its Petrov type ``settles down'' to 
Type D (assuming that one has ruled out other, non-Kerr, Type D spacetimes).

This line of reasoning was taken up in a paper by Campanelli, Lousto, 
and Zlochower~\cite{Campanelli2008}.  The central tool in their approach is a 
certain complex polynomial equation:
\beq
\Psi_4 \lambda^4 + 4 \Psi_3 \lambda^3 + 6 \Psi_2 \lambda^2 + 4 \Psi_1 \lambda + \Psi_0 = 0, \label{e:polynomial}
\eeq
the degeneracy of whose roots is known to correspond to the degeneracy of the 
principal null directions (assuming that $\Psi_4$ is nonzero).  The 
coefficients $\Psi_i$ of this polynomial are 
the so-called {\em Weyl scalars}, components (defined in 
Eq.~\eqref{e:WeylScalars} below) of the Weyl tensor in a given Newman-Penrose 
null tetrad.  If $\Psi_4$ is nonzero, then the fundamental theorem of algebra 
ensures that the polynomial has exactly four complex roots, counting 
multiplicities.

Once the four roots $\lambda_i$ have been computed for 
Eq.~\eqref{e:polynomial} at any 
given point, then one can also compute six distinct positive-definite root 
differences:
\beq \label{e:Deltaij}
\Delta_{ij} := | \lambda_i - \lambda_j |.
\eeq
If two of these root differences vanish and the other four are nonzero, 
meaning that the four roots coincide in two distinct pairs, then the spacetime 
is Type D at that 
point.  In Ref.~\cite{Campanelli2008} Campanelli {\em et al.}~took the next 
logical step: interpreting the two smallest $\Delta_{ij}$ values as 
measures of the ``nearness'' of an algebraically general (in that case, 
numerical) spacetime to Petrov Type D.  While this is a reasonable 
interpretation of $\Delta_{ij}$ and we will not suggest any fundamental 
modification to this approach of defining approximate algebraic speciality, 
there are important subtleties in this interpretation, not fully explored in 
Ref.~\cite{Campanelli2008}.  These subtleties relate to the geometrical 
meaning of $\Delta_{ij}$ and its behavior under 
tetrad transformations.  The main purpose of this paper is to explore these 
subtleties, present an alternative degeneracy measure that avoids certain 
blowups that are intricately related to the choice of tetrad (and should 
therefore not be considered physically relevant),
and apply both degeneracy measures to a numerical simulation 
from the {\tt SpEC} code~\cite{SpEC}.  In the process we will also 
investigate an interesting conclusion from Ref.~\cite{Campanelli2008}: that 
in the ringdown of a binary black hole merger to Kerr, the spacetime 
approaches Petrov Type II very quickly, and Type D much later.  We will argue 
that this conclusion is due essentially to a coordinate singularity on the 
space of null rays, and the fact that the tetrad used in 
Ref.~\cite{Campanelli2008} was much better suited to representing the 
degeneracy of one pair of principal null directions than the other pair, when 
the degeneracy measure $\Delta_{ij}$ is used.  The alternate degeneracy 
measure that 
we will introduce, $\Theta_{ij}$ defined in Eq.~\eqref{e:Thetadefined} below, 
shows both pairs of principal null directions approaching degeneracy at the 
same rate.

Though much of the discussion in this paper centers upon the behavior of these 
measures of degeneracy under tetrad transformations, we will unfortunately not 
be 
able to provide a measure of nearness to Petrov Type D that is fundamentally 
any more invariant than $\Delta_{ij}$.  This is because no such measure 
appears to exist.  Geometrically this fact can be understood in terms of the 
nonexistence of a boost-invariant geometry on the space of null rays in 
Minkowski space, an issue referred to physically as the ``relativistic 
aberration of starlight.''  This viewpoint is explored in more detail in 
Section~\ref{s:spinors} below.  

The issue can also be understood at the algebraic level, as in 
Petrov's original construction.  The problem shown in 
Eq.~\eqref{e:petrovproblem} can be written more compactly if one works in the 
three-complex-dimensional space of {\em anti-self-dual} bivectors rather than 
in the six-real-dimensional space of real bivectors.  In this space, the 
eigenbivector problem can be written as:
\beq
{W_{ab}}^{cd} Z_{cd} = \Lambda Z_{ab}, \label{e:asdeigenbivect}
\eeq
where $W_{abcd} := C_{abcd} + i \hspace{1mm}{}^\star C_{abcd}$, 
${}^\star Z_{ab} = - i Z_{ab}$, and $\Lambda$ is a complex number.  Because 
this is a three-dimensional problem one can expect three possible values 
for $\Lambda$, though the fact that the Weyl tensor is tracefree implies 
that these three eigenvalues must sum to zero.  The degeneracy of the 
eigenvalues and the completeness of 
the corresponding eigenspaces determine the classification of the Weyl 
tensor at the point under consideration.  If all three eigenvalues 
are distinct then the spacetime is algebraically 
general.  If two roots coincide, then the spacetime is either Type II or 
Type D.  If all three coincide (and therefore vanish, as they must sum to 
zero) then the spacetime is either Type III, Type N, or conformally flat.  The 
eigenvalues are geometrically defined at each point in spacetime, 
independent of the vector basis used to represent the eigenproblem.  The 
differences between these eigenvalues can therefore be used to construct 
invariant measures of the approach to algebraic speciality.  For example, the 
absolute value of the difference between the two nearest eigenvalues 
can be thought of as such an invariant measure.  Unfortunately this 
measure isn't very specific: it vanishes for 
Petrov Types II, D, III, and N.\footnote{In this sense it is like other scalar 
measures of algebraic speciality, such as the ``cross ratio'' of principal 
null directions, defined in Ref.~\cite{PenroseRindler1}, whose explicit 
relationship to the eigenvalues is described in Sec.~8.3 of 
Ref.~\cite{PenroseRindler2}, or the Baker-Campanelli ``speciality 
index''~\cite{BakerCampanelli2000}, which takes a special value for 
{\em any} type of algebraic speciality, but cannot distinguish between 
the various types.}  The latter pair can be distinguished from 
the former pair by the fact that all three eigenvalues vanish in Type III and 
Type N, but distinguishing Type II from Type D, or Type III from Type N, 
requires more information than just the eigenvalues.  

If two of the eigenvalues in Eq.~\eqref{e:asdeigenbivect} coincide, so that 
the eigenvalues can be written as $\{\Lambda, \Lambda, -2\Lambda\}$, then 
the distinction between Petrov Type II and Type D can be made by the following 
quantity\footnote{Incidentally, if one wishes to avoid the assumption that the 
spacetime is at least Type II, such that the eigenvalues can be written as 
$\{\Lambda, \Lambda, -2\Lambda\}$, this can be done with the help of certain 
curvature invariants.  See Ref.~\cite{FerrandoSaez2009}.}:
\beq
{T^{ab}}_{cd} := ( {W^{ab}}_{ef} - \Lambda I^{ab}_{ef} ) ( {W^{ef}}_{cd} + 2 \Lambda I^{ef}_{cd} ), \label{e:tensormeasure}
\eeq
where $I^{ab}_{cd}$ is the identity operator on the space of anti-self-dual 
bivectors.  The object ${T^{ab}}_{cd}$ vanishes in Type D, but not in 
Type II~\cite{ExactSolutionsBook}.  The difficulty with 
using this as a measure of nearness to Petrov Type D is that it is a tensorial 
object, and its components are, by definition, basis-dependent.  In order to 
collapse this object to a single number for each point in spacetime, one might 
hope to construct a positive-definite tensor norm:
\beq
Q := m_{ae} \hmm m_{bf} \hmm m^{cg} \hmm m^{dh} \hmm {T^{ab}}_{cd} \hmm {\ovl{T}^{ef}}{}_{gh},
\eeq
where $m_{ab}$ is a positive-definite inner product on spacetime.  
Unfortunately, the only inner product that one naturally has available on 
spacetime is the indefinite spacetime metric.  If a timelike 
``observer'' is
introduced, with unit tangent vector $u^a$, then one can construct a positive 
definite inner product as:
\beq
m_{ab} := g_{ab} + 2 u_a u_b,
\eeq
but then the quantity $Q$ is not strictly a scalar, as its definition is 
dependent on the extra structure of this observer.

Though the language is very different in the geometric approach involving 
principal null directions, we will find in Section~\ref{s:spinors} that the 
ambiguity in defining a measure of ``nearness'' to Petrov Type D is in that 
context essentially the same as here, requiring the choice of a 
timelike observer at every point in spacetime.  While this state of affairs 
seems to endanger any attempt at defining the nearness to any specific 
Petrov class, there are some cases where a well-defined fleet of 
observers can be chosen.  In particular, in any stationary spacetime, one can 
choose the stationary observers.  In cases such as the ringdown to Kerr 
geometry, one can expect an ``approximate'' stationarity to be approached 
at late times, again providing a preferred class of observers at least during 
the late ringdown.  A major practical goal of this paper will be to study 
this ringdown process, as in Ref.~\cite{Campanelli2008}.  In particular, we 
will argue that the degeneracy measure $\Delta_{ij}$ that was used in 
Ref.~\cite{Campanelli2008} is in some sense adapted to a {\em null} observer 
that happened in that case to be nearly aligned with one of the nearly 
degenerate pairs of principal null directions, making this pair of null 
directions seem much more degenerate, and the other, much less.  This 
causes the appearance of a holdup in Petrov Type II before the 
spacetime geometry falls to Type D.

The structure of this paper is as follows: in Sec.~\ref{s:tetraddependence} 
we will investigate the ambiguity of the measure $\Delta_{ij}$ under 
tetrad rotations, particularly those that leave the timelike tetrad leg 
fixed.  In Sec.~\ref{s:quasikinnersley} we will emphasize the fact that a 
tetrad well-suited to gravitational wave extraction, in particular the 
quasi-Kinnersley tetrad~\cite{Nerozzi2005}, may be particularly ill-suited to 
measuring the nearness to Petrov Type D using $\Delta_{ij}$.  In 
Sec.~\ref{s:spinors} we will describe the geometry underlying 
$\Delta_{ij}$ in spinorial language, and in the process motivate 
a modification that is much better suited to 
situations such as the ringdown to Kerr geometry.  In Sec.~\ref{s:results} 
we will present numerical results applying these degeneracy measures to 
a binary black hole merger simulation, demonstrating in detail the 
approach to Petrov Type D.  Finally in Sec.~\ref{s:discussion} we conclude 
with further discussion of the subtleties that have been addressed, and those 
that remain.

\section{Tetrad dependence}
\label{s:tetraddependence}

The method put forth in Ref.~\cite{Campanelli2008} to define nearness to a 
Petrov 
class begins with the polynomial in Eq.~\eqref{e:polynomial}, whose 
coefficients 
are components of the Weyl tensor in a Newman-Penrose tetrad~\cite{Newman1962}:
\begin{subequations}\label{e:WeylScalars}
\beqa
\Psi_0 &:=& C_{a b c d} \ell^a m^b \ell^c m^d,\\
\Psi_1 &:=& C_{a b c d} \ell^a n^b \ell^c m^d,\\
\Psi_2 &:=& \frac{1}{2} C_{a b c d} \left( \ell^a n^b \ell^c n^d - \ell^a n^b m^c \overline m^d \right),\\
\Psi_3 &:=& C_{a b c d} n^a \ell^b n^c \overline m^d,\\
\Psi_4 &:=& C_{a b c d} n^a \overline m^b n^c \overline m^d.
\eeqa
\end{subequations}
The tetrad $\{ \ell^a, n^a, m^a, \overline m^a \}$ is made up of two 
future-directed real null vectors $\ell^a$ and $n^a$ and two complex 
conjugate null vectors $m^a$ and $\overline m^a$ with spacelike real and 
imaginary parts.  These vectors are normalized by the conditions:
\beqa
\ell_a n^a &=& -1,\\
m_a \ovl m^a &=& 1,\\
\ell_a m^a = n_a m^a &=& 0.
\eeqa

These normalization conditions are preserved by three types of tetrad 
transformations which, taken together, are equivalent to the proper Lorentz 
group.  First, there are the ``null rotations about $\ell^a$,'' sometimes 
referred to as the ``Type I'' transformations\footnote{To avoid confusion 
with the Petrov types, we will hereafter refer to tetrad transformations as 
``null rotations about $\ell^a$,'' ``null rotations about $n^a$,'' or 
``spin boosts,'' rather than ``Type I,'' ``Type II,'' or ``Type III.''}:
\begin{subequations}\label{e:nullrot_l}
\beqa
\ell^a &\mapsto& \ell^a,\\
m^a &\mapsto& m^a + a \ell^a,\\
\ovl m^a &\mapsto& \ovl m^a + \ovl a \ell^a,\\
n^a &\mapsto& n^a + \ovl a m^a + a \ovl m^a + a \ovl a \ell^a,
\eeqa
\end{subequations}
where $a$ is a complex number, and can vary over spacetime.  
Second, there are the null rotations about $n^a$, sometimes referred to as 
``Type II'' transformations:
\begin{subequations}\label{e:nullrot_n}
\beqa
\ell^a &\mapsto& \ell^a + \ovl b m^a + b \ovl m^a + b \ovl b n^a,\\
m^a &\mapsto& m^a + b n^a,\\
\ovl m^a &\mapsto& \ovl m^a + \ovl b n^a,\\
n^a &\mapsto& n^a,
\eeqa
\end{subequations}
for complex $b$.  Third, there are the ``spin-boost'' transformations, 
sometimes referred to as the ``Type III'' transformations:
\begin{subequations}\label{e:spinboost}
\beqa
\ell^a &\mapsto& |c|^2 \ell^a,\\
m^a &\mapsto& e^{2 i \arg(c)} m^a,\\
\ovl m^a &\mapsto& e^{- 2 i \arg(c)} \ovl m^a,\\
n^a &\mapsto& |c|^{-2} n^a,
\eeqa
\end{subequations}
for complex $c$.  

These transformation laws for the tetrad imply transformation laws for the 
Weyl scalars.  Under the null rotations about $\ell^a$, 
Eqs.~\eqref{e:nullrot_l}, the Weyl scalars transform as:
\begin{subequations}\label{e:psi_nr_l}
\beqa
\Psi_0 &\mapsto& \Psi_0,\\
\Psi_1 &\mapsto& \Psi_1 + \ovl a \Psi_0,\\
\Psi_2 &\mapsto& \Psi_2 + 2 \ovl a \Psi_1 + \ovl a^2 \Psi_0,\\
\Psi_3 &\mapsto& \Psi_3 + 3 \ovl a \Psi_2 + 3 \ovl a^2 \Psi_1 + \ovl a^3 \Psi_0,\\
\Psi_4 &\mapsto& \Psi_4 + 4 \ovl a \Psi_3 + 6 \ovl a^2 \Psi_2 + 4 \ovl a^3 \Psi_1 + \ovl a^4 \Psi_0.
\eeqa
\end{subequations}
Under null rotations about $n^a$, Eqs.~\eqref{e:nullrot_n}, the 
Weyl scalars transform as:
\begin{subequations}\label{e:psi_nr_n}
\beqa
\Psi_0 &\mapsto& b^4 \Psi_4 + 4 b^3 \Psi_3 + 6 b^2 \Psi_2 + 4 b \Psi_1 + \Psi_0,\\
\Psi_1 &\mapsto& b^3 \Psi_4 + 3 b^2 \Psi_3 + 3 b \Psi_2 + \Psi_1,\\
\Psi_2 &\mapsto& b^2 \Psi_4 + 2 b \Psi_3 + \Psi_2,\\
\Psi_3 &\mapsto& b \Psi_4 + \Psi_3,\\
\Psi_4 &\mapsto& \Psi_4.
\eeqa
\end{subequations}
Finally, under the spin boosts, Eqs.~\eqref{e:spinboost}, the Weyl scalars 
simply rescale, as:
\beq \label{e:psi_sb}
\Psi_n \mapsto c^{2(2-n)} \Psi_n.
\eeq

The transformation laws for the coefficients of the polynomial in 
Eq.~\eqref{e:polynomial} imply transformation laws for the roots.  It is 
straightforward to show that under the transformation in 
Eq.~\eqref{e:psi_nr_l}, the roots of the polynomial transform as:
\beq
\lambda \mapsto \frac{\lambda}{\ovl a \lambda + 1}. \label{e:lambda_nr_l}
\eeq
Under transformations of the form~\eqref{e:psi_nr_n}, 
the roots transform as:
\beq
\lambda \mapsto \lambda + b. \label{e:lambda_nr_n}
\eeq
Finally, under spin-boost transformations, Eq.~\eqref{e:psi_sb}, the roots 
transform as:
\beq
\lambda \mapsto c^2 \lambda. \label{e:lambda_sb}
\eeq

In Ref.~\cite{Campanelli2008}, nearness to Petrov Type D was mainly argued 
through the approach of the absolute values of root differences ($\Delta_{ij}$ 
as defined in Eq.~\eqref{e:Deltaij})
to zero.  While this quantity would indeed be 
expected to vanish when $\lambda_i$ and $\lambda_j$ constitute a degenerate 
root pair, if they are not exactly degenerate, then the foregoing discussion 
implies that this difference is not invariant under tetrad 
transformations.  The transformation in Eq.~\eqref{e:lambda_nr_n} would leave 
$\Delta_{ij}$ unchanged, but that in Eq.~\eqref{e:lambda_sb} would 
directly rescale any given root difference (though the complex phase of $c$ 
would not appear in the absolute value), and transformations of the 
form~\eqref{e:lambda_nr_l} would change $\Delta_{ij}$ in a more 
complicated way.  Arbitrary Lorentz transformations, given by arbitrary 
combinations of the above transformations, could alter 
$|\lambda_i - \lambda_j|$ in a very complicated manner.  

To investigate the practical relevance of this tetrad ambiguity in the 
degeneracy measure $\Delta_{ij}$, let us consider a particular case of 
possible physical relevance that requires a combination of all three of the 
above tetrad transformations.  Take the case where one has 
a particular timelike vector defined at a point in spacetime, for example a 
timelike Killing vector, or a kind of approximate Killing vector 
generating time translations in a spacetime that is approaching stationarity 
in some sense.  Given a Newman-Penrose tetrad $\{ \ell^a, n^a, m^a, \ovl m^a \}$, one can construct a standard orthonormal tetrad in the following way:
\begin{subequations}\label{e:orthonormal}
\beqa
e_0^a &:=& (\ell^a + n^a)/\sqrt{2},\\
e_1^a &:=& \sqrt{2} \hsp{\rm Re}\left[ m^a \right],\\
e_2^a &:=& \sqrt{2} \hsp {\rm Im}\left[ m^a \right],\\
e_3^a &:=& (\ell^a - n^a)/\sqrt{2}.
\eeqa
\end{subequations}
Rotations of the tetrad legs in the $\vec e_1$--$\vec e_2$ plane are easily 
accomplished, through a simple spin-boost transformation with the parameter 
$c = e^{i \Phi/2}$.  Such rotations also however leave the degeneracy 
measure $\Delta_{ij}$ unchanged.  For a nontrivial test case, consider a 
rotation in the $\vec e_1$--$\vec e_3$ plane:
\begin{subequations}\label{e:xzrotation}
\beqa
\vec e_0{}^\prime &=& \vec e_0\\
\vec e_1{}^\prime &=& \cos(\Phi) \vec e_1 - \sin(\Phi) \vec e_3\\
\vec e_2{}^\prime &=& \vec e_2\\
\vec e_3{}^\prime &=& \cos(\Phi) \vec e_3 + \sin(\Phi) \vec e_1
\eeqa
\end{subequations}
A straightforward calculation shows that such a transformation 
can be carried out by a sequence of the above transformations.  First, one 
makes a null rotation about $\ell^a$, Eqs.~\eqref{e:nullrot_l}, with 
parameter $a = - \tan(\Phi/2)$.  Second, there is a null rotation about 
$n^a$, Eqs.~\eqref{e:nullrot_n}, with parameter $b = (1/2) \sin(\Phi)$.  
The final step is a spin boost, Eqs.~\eqref{e:spinboost}, with parameter 
$c = \sec(\Phi/2)$.  In this particular case, all three parameters are real.  
Composing the transformation laws for the roots, 
Eqs.~\eqref{e:lambda_nr_l}~--~\eqref{e:lambda_sb}, with these parameters, 
the resulting transformation law is:
\beq
\lambda^\prime = \frac{\lambda \cos(\Phi/2) - \sin(\Phi/2)}{\lambda \sin(\Phi/2) + \cos(\Phi/2)}.\label{e:lambda_spatrot}
\eeq
If we express $\lambda$ as a ratio of two complex numbers, 
$\lambda = \xi/\eta$, then Eq.~\eqref{e:lambda_spatrot} takes a very 
simple matrix form:
\beq
\begin{pmatrix} \xi^\prime \\ \eta^\prime \end{pmatrix} = \begin{bmatrix} \cos(\Phi/2) & - \sin(\Phi/2) \\ \sin(\Phi/2) & \cos(\Phi/2) \end{bmatrix} \begin{pmatrix} \xi \\ \eta \end{pmatrix}.\label{e:sl2c_spatrot}
\eeq
The general form of this matrix, for arbitrary reorientations using three 
Euler angles, is given in Eq.~(1.2.34) of Ref.~\cite{PenroseRindler1}.
The $SL(2,\mathbb{C})$ form of this transformation suggests a 
spinorial interpretation of $\lambda$, a point to which we will return in 
Sec.~\ref{s:spinors}.  

For now let us consider the behavior of the degeneracy measure $\Delta_{ij}$ 
under these spatial rotations.  For concreteness, consider the case 
where the four roots of Eq.~\eqref{e:polynomial} are 
$\lambda_1 = .005 + .047 i$, $\lambda_2 = .005 + .05 i$, 
$\lambda_3 = -5 + 15 i$, and $\lambda_4 = -5 + 15.5 i$.  These values are 
chosen to very roughly mirror the late-term values seen in Fig.~8 of 
Ref.~\cite{Campanelli2008}, with degeneracies roughly similar to those seen in 
Figs.~3 and 4 of that paper.  The values estimated here are extremely 
rough, and should not be taken as having any quantitative importance, but 
merely as tools for illustrating the qualitative features of the 
transformation law in Eq.~\eqref{e:lambda_spatrot}.  So long as one pair of 
nearly-degenerate roots is larger, by a few orders of magnitude, than the 
other pair, the qualitative behavior that we will describe seems roughly 
the same regardless of the particular choice of roots.

\begin{figure}
\begin{center}
\includegraphics[scale=.75]{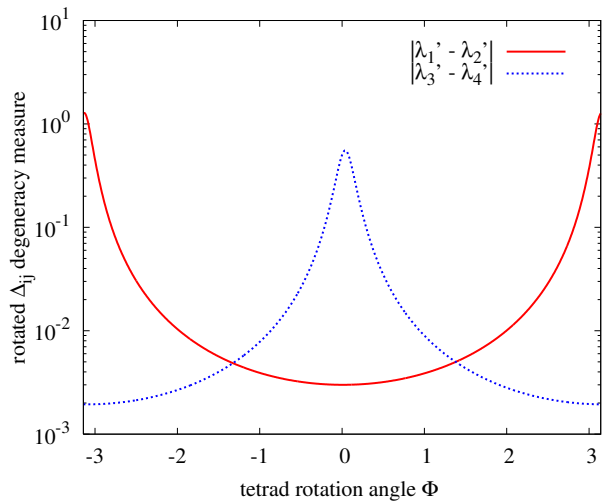}
\end{center}
\caption{ \label{f:analyticspatrot} Behavior of the degeneracy measure 
$\Delta_{ij}$ under the tetrad rotation in Eq.\eqref{e:xzrotation} for 
a particular (though essentially arbitrary) choice of roots, stated in the 
text.  Under a rotation through 180 degrees, the root pair that originally 
seemed more degenerate becomes less degenerate, and the pair that originally 
seemed less degenerate becomes more degenerate.  
}
\end{figure}

The degeneracy measure $\Delta_{ij}$ for the two most nearly degenerate root 
pairs, under rotation of the $\vec e_1$--$\vec e_3$ plane, is shown in 
Fig.~\ref{f:analyticspatrot}.  If the 
tetrad's spatial legs were rotated through about ninety degrees, then both 
root pairs would appear equally close to degeneracy.  If 
the tetrad were rotated through 180 degrees, then the root pair that 
originally appeared closer to degeneracy would begin to seem farther 
away from it, and the one that originally seemed less degenerate would seem 
more so.  

This variation in the degeneracy measure can be interpreted as a coordinate 
effect.  The quantity $\lambda$ has no inherent geometrical meaning without 
a particular reference tetrad.  It is essentially a coordinate on the space 
of null rays at a point in spacetime.  This space of null rays is 
topologically a two-dimensional sphere, as can be demonstrated by cutting 
a future null cone with a spacelike hyperplane, such as the $t=1$ plane in 
Minkowski space.  A two-sphere cannot be covered smoothly with a single 
coordinate patch.  If the quantity $\lambda$ is taken as a (complex) 
coordinate labeling all the null rays at a point, then there must be a 
coordinate singularity somewhere, near which coordinate distances are 
particularly ill-suited to representing the true geometry that may be defined 
on the manifold.  We will study this issue in more detail in 
Sec.~\ref{s:spinors}.  For now we simply note that the locations of the 
sharp peaks in 
Fig.~\ref{f:analyticspatrot} seem to imply that such a coordinate singularity 
may have a particularly strong effect in the original, unrotated tetrad.  
The following section gives an extreme example of this effect.  

\section{The quasi-Kinnersley tetrad:}
\label{s:quasikinnersley}

It appears from the results of the previous section that a tetrad that seems 
reasonable for purposes of wave extraction can be particularly ill-suited 
to the problem of defining nearness to a Petrov class.  To investigate 
this point in more detail, here we consider a special family of tetrads 
designed especially for wave extraction.  

Consider an algebraically general spacetime (eventually we will allow this 
spacetime to ``asymptote'' toward Petrov Type D, but we will consider it 
always to be, strictly speaking, Type I).  As described in 
Ref.~\cite{Nerozzi2005}, 
at any point where the Weyl tensor is Type I, there are 
precisely three distinct families of tetrads in which two particular Weyl 
scalars vanish, $\Psi_1 = \Psi_3 = 0$ (they each amount to {\em families} of 
tetrads, rather than three particular tetrads, because this condition is 
preserved by the spin-boost freedom).  
A particular tetrad field, chosen from these three families to coincide with 
the conventional Kinnersley tetrad near infinity, is often referred to as a 
{\em quasi-Kinnersley tetrad}.  The usual purpose of such a tetrad is to aid 
in gravitational wave extraction, where the relative uniqueness of the 
tetrad provides a preferred reference frame in which to define gravitational 
radiation.  Such a tetrad also simplifies the polynomial in 
Eq.~\eqref{e:polynomial}:
\beq
\Psi_4 \lambda^4 + 6 \Psi_2 \lambda^2 + \Psi_0 = 0.\label{e:qKpoly}
\eeq
If, as we are assuming, the spacetime is strictly Type I, and not a more 
special algebraic type, then $\Psi_4$ and $\Psi_0$ will be nonzero.  In the 
limit that the spacetime asymptotes to Type D, they will both settle to zero, 
indicating a failure of the polynomial roots to represent the principal null 
directions in the conventional sense.  What we wish to 
investigate is the behavior of these roots as this limit is approached.  

Carrying on under the assumption that $\Psi_4$ is nonzero, the roots of 
Eq.~\eqref{e:qKpoly} are readily found.
%\beqa
%\lambda^2 &=& -\frac{3 \Psi_2}{\Psi_4} \pm \sqrt{\left(\frac{3 \Psi_2}{\Psi_4}\right)^2 - \frac{\Psi_0}{\Psi_4}},\\
%&=& \frac{3 \Psi_2}{\Psi_4} \left( -1 \pm \sqrt{1 - \frac{\Psi_0 \Psi_4}{9 \Psi_2^2}}\right).
%\eeqa
\beq
\lambda^2 = \frac{3 \Psi_2}{\Psi_4} \left( -1 \pm \sqrt{1 - \frac{\Psi_0 \Psi_4}{9 \Psi_2^2}}\right).
\eeq

If we now consider the approach to a Kerr geometry, in which the 
quantity $\Psi_0 \Psi_4/(9 \Psi_2^2)$ approaches zero, we can expand the 
square root in the above expression to first order in this small 
quantity\footnote{Note that the numerator in this quantity, $\Psi_0 \Psi_4$, 
which we are evaluating in a ``transverse frame'' --- one where 
$\Psi_1 = \Psi_3 = 0$ --- is the Beetle-Burko ``radiation scalar'' 
described in Ref.~\cite{BeetleBurko2002}.}:
\beqa
\lambda^2 &\approx& \frac{3 \Psi_2}{\Psi_4} \left[-1 \pm \left( 1 - \frac{\Psi_0 \Psi_4}{18 \Psi_2^2}\right)\right],\\
\lambda &\approx& \left\{ \pm \sqrt{-\frac{6 \Psi_2}{\Psi_4}}, \pm \sqrt{-\frac{\Psi_0}{6 \Psi_2}} \right\}. \label{e:halfrate}
\eeqa
So in the Kerr limit, as $\Psi_4 \rightarrow 0$ and $\Psi_0 \rightarrow 0$, 
two of these roots approach zero, and so does their difference, 
but the other two approach infinity (this is a standard behavior of 
polynomial roots as the leading polynomial coefficient approaches zero).  
Moreover, they approach the point at infinity from different directions, so 
their difference also approaches infinity.  Geometrically, one would think 
that the problem is solved if the roots are considered not as numbers on the 
complex plane, but as points on the Riemann sphere.  The roots that blow up 
would then be taken as approaching a degenerate root at the point at 
infinity.  In the 
following section, we will motivate such a viewpoint in detail, and in the 
process, outline the geometrical meaning of the degeneracy measure 
$\Delta_{ij}$ and present an alternative that avoids the danger of 
representing any particular null ray as a ``point at infinity.''  

Before moving on, though, we should investigate the robustness of this 
behavior under 
tetrad rotations.  In practice, the tetrads used in numerical relativity 
simulations are 
usually simple coordinate-adapted tetrads, rather than carefully constructed 
quasi-Kinnersley tetrads.  But because they are usually adapted to a spherical 
coordinate basis, they very roughly tend to approximate the quasi-Kinnersley 
tetrad during black hole ringdown, by force of topology alone.  For this 
reason, it is interesting to investigate the behavior of the polynomial roots 
not only in the quasi-Kinnersley tetrad, but also in tetrads slightly offset 
from it.  

\begin{figure}
\begin{center}
\includegraphics[scale=.75]{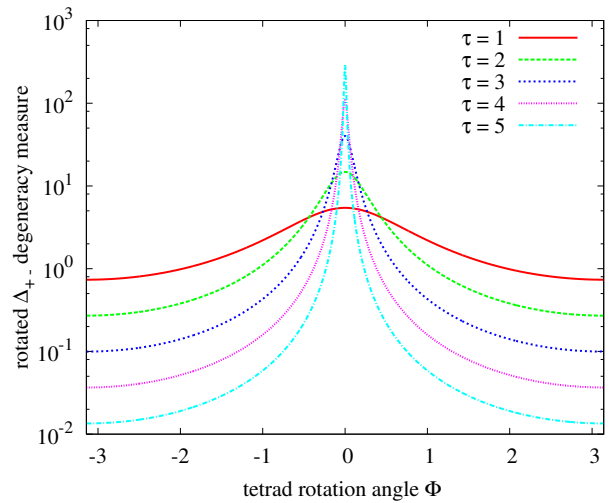}
\end{center}
\caption{ \label{f:qKoffset_manytimes} Profiles of the behavior of the 
degeneracy measure $\Delta_{ij}$ under tetrad rotations of the form in 
Eqs.~\eqref{e:xzrotation} from a quasi-Kinnersley tetrad, for various values 
of a fiducial time coordinate, assuming that this degeneracy measure grows 
exponentially in this fiducial time coordinate for the true quasi-Kinnersley 
tetrad ($\Phi = 0$).  The peak value grows exponentially in time, by 
construction, but values well outside the peak decay exponentially in time.  
The peak sharpens as it grows, so that values slightly offset from the peak 
grow initially, and decay later.}
\end{figure}

\begin{figure}
\begin{center}
\includegraphics[scale=.75]{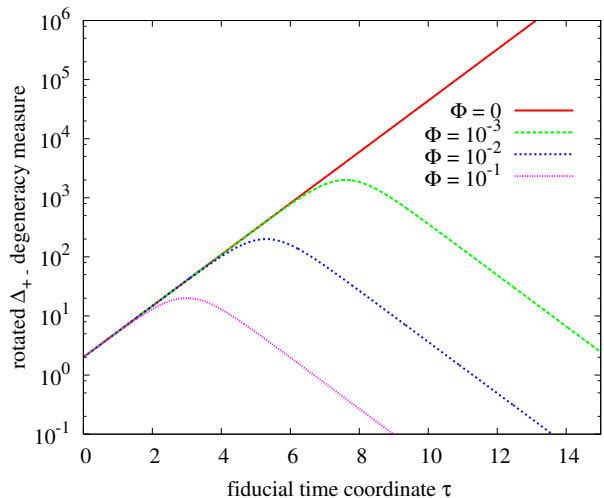}
\end{center}
\caption{ \label{f:qKoffset_manyangles} Behavior over time of the degeneracy 
measure $\Delta^\prime_{+-}(\Phi) := |\lambda^\prime_+ - \lambda^\prime_-|$, 
for roots $\lambda^\prime_\pm$ given by Eq.~\eqref{e:ringdown_spatrot}, for a 
few values of the tetrad rotation parameter $\Phi$.  Each curve initially 
grows exponentially in the time parameter $\tau$ before eventually falling.  
The more offset the tetrad is from the quasi-Kinnersley tetrad ($\Phi = 0$), 
the sooner this turnaround occurs.}
\end{figure}

In particular, consider the ringdown to a Kerr black hole, where in the true 
Kinnersley tetrad of a Kerr background one would expect the absolute value 
of $\Psi_4$ to approach zero 
exponentially in time at a rate determined by the quasinormal frequencies of 
the hole.  The roots $\pm \sqrt{-6 \Psi_2/\Psi_4}$ would then be expected to 
grow exponentially at half that rate.  Consider, for example, a case where 
$\sqrt{-6 \Psi_2/\Psi_4} = i \exp(\tau)$ for some time function 
$\tau$.\footnote{The imaginary factor $i$ is inserted to avoid the rotated 
tetrad vector exactly coinciding with a principal null direction at some time, 
a possibility that, however possible, would not be expected to occur 
generically.}  The roots $\pm i \exp(\tau)$, if the tetrad were rotated 
spatially as in Eq.~\eqref{e:lambda_spatrot}, would instead take the values:
\beq
\lambda^\prime_\pm = \frac{\pm i \exp(\tau) \cos(\Phi/2) - \sin(\Phi/2)}{\pm i \exp(\tau) \sin(\Phi/2) + \cos(\Phi/2)}.\label{e:ringdown_spatrot}
\eeq
So in the limit that $\tau \rightarrow \infty$, these roots would become 
degenerate at the value $\cot(\Phi/2)$, and their difference, as measured 
by $\Delta_{ij}$, would eventually fall to zero.  The details of how this 
occurs are plotted in Figs.~\ref{f:qKoffset_manytimes} 
and~\ref{f:qKoffset_manyangles}.

In Fig.~\ref{f:qKoffset_manytimes},
the profile of 
$\Delta^\prime_{+-} := |\lambda^\prime_+ - \lambda^\prime_-|$, as a function 
of tetrad rotation angle $\Phi$ in Eq.~\eqref{e:ringdown_spatrot}, is shown 
for a few values of the fiducial time label $\tau$.  Each curve is peaked, as 
in Fig.~\ref{f:analyticspatrot}, at the quasi-Kinnersley tetrad.  The value of 
this peak grows exponentially in $\tau$, while the values well outside the 
peak (representing more arbitrary tetrads) decay exponentially in $\tau$.  
What is of particular interest to us is the behavior {\em near} $\Phi = 0$.  
Because the peak sharpens as it grows, values of $\Delta^\prime_{+-}$ slightly 
offset 
from $\Phi = 0$ grow initially, and eventually decay.  This behavior is more 
clearly visible in Fig.~\ref{f:qKoffset_manyangles},
where $\Delta^\prime_{+-}$ is 
shown as a function of $\tau$ for various choices of the offset angle.  For 
any fixed nonzero value of $\Phi$, the curve initially grows exponentially 
before eventually falling at the same rate.  
The smaller this rotation angle is, the later the 
curve turns around.  So the nearer a tetrad is to 
quasi-Kinnersley, the longer it takes for $\Delta^\prime_{+-}$ to eventually 
decay as one might naively expect.

Incidentally, we should note that the ``peak'' in 
Fig.~\ref{f:qKoffset_manytimes}, and indeed in Fig.~\ref{f:analyticspatrot}, 
is actually a saddle point if considered in a larger space of tetrad 
rotations.  To keep the discussion simple, in 
Sec.~\ref{s:tetraddependence} we considered only rotations in the local 
$\vec e_1$--$\vec e_3$ tangent plane.  Had we considered the case of general 
rotations of the spatial tetrad, as in Eq.~(1.2.34) of 
Ref.~\cite{PenroseRindler1}, we would have found that the degeneracy measure 
$\Delta_{ij}$ becomes infinite whenever the rotated tetrad $\vec n$ vector 
coincides with a principal null direction.

\section{Interpreting degeneracy measures:}
\label{s:spinors}

The geometrical underpinnings of the polynomial in Eq.~\eqref{e:polynomial}, 
and the sense in which $\lambda$ constitutes a coordinate on the space of 
null rays, 
are most cleanly explained in the language of two-component spinors.  Because 
many numerical relativists are unfamiliar with this formalism I will attempt 
to 
keep the discussion self-contained by briefly reviewing crucial elements as 
we go along.  For a detailed account of spinor methods in spacetime geometry, 
see Refs.~\cite{PenroseRindler1, PenroseRindler2}, or for a more compact 
treatment specifically geared to numerical relativists, see 
Ref.~\cite{StewartBook}.

Throughout this paper, objects with capital latin indices will be referred to 
as spinors, elements of a two-complex-dimensional vector space (or its higher 
tensorial orders).  The complex 
conjugate of a spinor is also a spinor, but is defined in a {\em different} 
spinor space, because complex conjugation does not commute with multiplication 
by a complex scalar.  To distinguish objects in spinor space from objects in 
the complex conjugate space, we will apply the standard convention of 
appending indices referring to the latter space with a prime:
\beq
\overline{\alpha^A} = \overline \alpha^{A'}.
\eeq

Spinors are useful in relativity theory because a simple 
correspondence exists between spinor space and Minkowski space (and therefore 
also to the tangent space to spacetime at any given point, given an 
orthonormal tetrad).  From a spinor $\alpha^A$ a unique vector can be 
constructed in Minkowski space:
\beq
V^a = \alpha^A \overline \alpha^{A'} {\sigma^a}_{AA'}, \label{e:correspondence}
\eeq
where the ${\sigma^a}_{AA'}$ are soldering forms, specifically referred to as 
{\em Infeld--van den Waerden symbols}, conventionally represented as Pauli 
matrices.  In practice, the transformation provided by ${\sigma^a}_{AA'}$ is 
often (and hereafter) taken as implied, with pairs of capital latin indices 
(one primed and one unprimed, with the same letter) taken to correspond 
abstractly to a single spacetime index.  

Vectors defined directly from univalent spinors as in 
Eq.~\eqref{e:correspondence} turn out always to be null.  For that reason 
univalent spinors can be understood as defining null vectors in spacetime.  
The standard geometrical interpretation of a univalent spinor (again, see 
Ref.~\cite{PenroseRindler1, StewartBook}) is as a ``null flag,'' a null vector 
with a particular spacelike half-plane attached to it.  This flag plane, 
encoded in the spinor's complex phase, is unimportant for our current 
purposes.  

The spacetime Weyl tensor can be written in terms of a four-index 
totally symmetric object called the {\em Weyl spinor} $\Psi_{ABCD}$ and 
the antisymmetric metric $\epsilon_{AB}$ on spinor space:
\beq
W_{abcd} = \Psi_{ABCD} \overline\epsilon_{A'B'} \overline\epsilon_{C'D'}.
\eeq
Here, as in the introduction, $W_{abcd}$ refers to the Weyl tensor in its 
complex anti-self-dual 
form, $W_{abcd} := C_{abcd} + i \hspace{1mm}{}^\star C_{abcd}$.  A basic 
result in spinor algebra (due essentially to the fundamental theorem of 
algebra) is that any totally symmetric spin tensor can be decomposed into 
a symmetrized product of univalent spinors.  In particular, for the Weyl 
spinor,
\beq
\Psi_{ABCD} = \alpha_{(A} \beta_B \gamma_C \delta_{D)},\label{e:principaldecomp}
\eeq
for univalent spinors $\alpha_A$, $\beta_B$, $\gamma_C$, $\delta_D$ defined 
up to arbitrary complex scaling (any one of them can be scaled at the cost of 
inversely scaling another).  These are referred to as {\em principal 
spinors} of $\Psi_{ABCD}$.  Because 
the principal spinors are defined only up to a complex scaling, their 
corresponding null vectors are defined only up to an arbitrary real scaling, 
and their flag planes are completely undefined.  The corresponding null 
vectors, defined up to scale, are the principal null directions 
of the Weyl tensor at the spacetime point under consideration.    

Because the metric on spinor space, $\epsilon_{AB}$, is antisymmetric, all 
spinors have vanishing norm.
\beq
\alpha_A \alpha^A = 0.
\eeq
For this reason, the condition for $\alpha^A$ to be a principal 
spinor of the Weyl spinor is:
\beq
\Psi_{ABCD} \alpha^A \alpha^B \alpha^C \alpha^D = 0.\label{e:prinspinor}
\eeq
To consider this equation more concretely, we introduce a basis, 
$\{o^A, \iota^A\}$, in spinor space, normalized by the standard condition 
$\epsilon_{AB} o^A \iota^B = 1$.  Such a spin dyad is 
equivalent\footnote{Strictly speaking, the correspondence is two-to-one, as 
the spin dyad $\{-o^A, -\iota^A\}$ defines the same tetrad as 
$\{o^A, \iota^A\}$.  The distinction, however, is not important here.} to a 
Newman-Penrose tetrad through the definitions 
$\ell^a = o^A \ovl o^{A'}$, $n^a = \iota^A \ovl \iota^{A'}$, 
$m^a = o^A \ovl \iota^{A'}$.
Given a spin dyad, an arbitrary spin vector can be written as:
\beq
\alpha^A = \eta o^A + \xi \iota^A,
\eeq
for complex components $\eta$, $\xi$.  Because we are only interested in 
spinors up to arbitrary complex scaling, we can divide by $\eta$ to let:
\beq
\alpha^A = o^A + \zeta \iota^A,\label{e:zetaintroduced}
\eeq
where $\zeta = \xi/\eta$ is a possibly-infinite complex number, an element of 
the one-point-compactified complex 
plane, ${\mathbb C} \cup \{\infty\}$, the Riemann sphere.

Scaling an arbitrary spinor to be of this form, and inserting it into 
Eq.~\eqref{e:prinspinor}, the resulting equation is:
\beq
\Psi_4 \zeta^4 + 4 \Psi_3 \zeta^3 + 6 \Psi_2 \zeta^2 + 4 \Psi_1 \zeta + \Psi_0 = 0,\label{e:polynomial_zeta}
\eeq
where we have used the standard spinorial definition of the Weyl scalars:
\begin{subequations}
\beqa
\Psi_0 &:=& \Psi_{ABCD} o^A o^B o^C o^D,\\
\Psi_1 &:=& \Psi_{ABCD} o^A o^B o^C \iota^D,\\
\Psi_2 &:=& \Psi_{ABCD} o^A o^B \iota^C \iota^D,\\
\Psi_3 &:=& \Psi_{ABCD} o^A \iota^B \iota^C \iota^D,\\
\Psi_4 &:=& \Psi_{ABCD} \iota^A \iota^B \iota^C \iota^D.
\eeqa
\end{subequations}
We thus find, comparing Eq.~\eqref{e:polynomial} with 
Eq.~\eqref{e:polynomial_zeta}, that the quantity $\lambda$ can be interpreted 
as the complex stereographic coordinate $\zeta$ on the Riemann 
sphere, and in particular, as defining a spinor $\alpha^A$ of the form in 
Eq.~\eqref{e:zetaintroduced} in a given spin dyad.  Hereafter we will 
consider $\zeta$ and $\lambda$ to be the same quantity, and use the symbols 
interchangably.  

This stereographic interpretation of $\zeta$ (or $\lambda$) 
is not merely a formality.  As 
described in Chapter 1 of Ref.~\cite{PenroseRindler1}, the space of 
future-directed null rays at a point in spacetime is topologically a 
two-sphere.  This can be demonstrated by cutting a future null cone with a 
spacelike 3-plane, given by $t=1$ in some local Minkowski coordinate system.  
Furthermore, if we choose a particular such cut, whose intersection with the 
null cone we will label $S^+$ and call the {\em anti-celestial sphere}, 
after Ref.~\cite{PenroseRindler1}, the metric 
induced on this two-sphere from that in the local Minkowski spacetime is:
\beq
ds^2 = \frac{4 d\zeta d\ovl \zeta}{(1 + \zeta \ovl \zeta)^2},\label{e:unitspherestereo}
\eeq
where for the coordinate we have chosen the $\zeta$ value of the spinor, of 
the form in Eq.~\eqref{e:zetaintroduced}, whose associated null direction 
intersects $S^+$.  Applying the transformation to conventional 
spherical coordinates,
\beq
\zeta = e^{i \phi} \cot(\theta/2),\label{e:stereographic}
\eeq
we arrive at the standard form of the unit sphere metric:
\beq
ds^2 = d\theta^2 + \sin^2(\theta) d\phi^2.\label{e:unitspheresphere}
\eeq

As a geometrical method for defining the nearness of two null directions 
to degeneracy, one can consider the metric~\eqref{e:unitspheresphere} on 
the anti-celestial sphere.  If $\zeta_i$ and $\zeta_j$ are two roots 
of Eq.~\eqref{e:polynomial}, then one can translate them to spherical 
coordinates $(\theta_i, \phi_i)$, $(\theta_j, \phi_j)$ by inverting 
Eq.~\eqref{e:stereographic}, and then use the metric distance function 
on the unit sphere, given by the haversine formula as:
\beq
\begin{split}
\Theta_{ij} := &2 \arcsin \left\{\left(\sin^2\left[\left(\theta_i-\theta_j\right)/2\right]\right.\right.\\
& \left. \left.+ \sin \theta_i \sin \theta_j \sin^2\left[(\phi_i-\phi_j)/2\right]\right)^{1/2}\right\}.
\end{split}
\eeq
This can also be written directly in terms of the stereographic coordinates as:
\beq
\Theta_{ij} = 2 \arcsin \left[ \frac{|\zeta_i - \zeta_j|}{\sqrt{(1+\zeta_i \ovl \zeta_i)(1+\zeta_j \ovl \zeta_j)}}\right]. \label{e:Thetadefined}
\eeq
As one can verify by a direct substitution of Eq.~\eqref{e:lambda_spatrot}, 
this degeneracy measure is invariant under spatial rotations of the form 
in Eq.~\eqref{e:xzrotation}, or indeed any tetrad rotation that leaves the 
timelike tetrad leg invariant.

We must stress, however, that even this is not a totally invariant measure 
of degeneracy.  In fact, there are fundamentally as many degrees of ambiguity 
in this measure as there are in $|\zeta_i - \zeta_j|$.  The ambiguity in 
$\Theta_{ij}$ is encoded in the choice of cut one makes to the null cone in 
order to construct $S^+$.  This can be interpreted physically as a 
result of the special relativistic effect known as ``relativistic aberration 
of starlight,'' by which the inferred geometry of the celestial (or as in 
this case, anti-celestial) sphere is conformally mapped under Lorentz 
boosts.  

A geometrical interpretation of the degeneracy measure 
$\Delta_{ij} := |\zeta_i - \zeta_j|$ can be found in spinor space.  
Two spinors $\alpha_1^A$ and $\alpha_2^A$ are proportional --- and therefore 
their associated real null vectors are proportional --- if and only if  
their antisymmetrized product vanishes:
\beq
\brak{\alpha_1}{\alpha_2} := \epsilon_{AB} \alpha_1^A \alpha_2^B = 0.
\eeq
It is tempting to use this quantity as a 
measure of the degeneracy of the null rays associated with $\alpha_1^A$ and 
$\alpha_2^A$, but we must remember to account for the scaling ambiguity of 
the spinors.  If $\brak{\alpha_1}{\alpha_2}$ is nonzero, then an arbitrary 
rescaling of either spinor, which should not alter any reasonable 
measure of the degeneracy of the null rays, would directly rescale 
$\brak{\alpha_1}{\alpha_2}$.  This ambiguity must be fixed by imposing a 
condition on the scaling of $\alpha_1^A$ and $\alpha_2^A$.  One possibility, 
given a particular spin dyad $\left\{ o^A, \iota^A \right\}$, is to assume 
that the spinors are of the form~\eqref{e:zetaintroduced}, with 
$\alpha_1^A = o^A + \zeta_1 \iota^A$, and 
$\alpha_2^A = o^A + \zeta_2 \iota^A$.  This condition 
can be stated for the associated null vectors $V_1^a$ and $V_2^a$ in 
terms of the Newman-Penrose tetrad as:
\beq
V_i^a n_a = - 1, \label{e:nullcut}
\eeq
for $i \in \{1, 2\}$, along with the conditions that the $V_i^a$ are real null 
vectors, and where the Newman-Penrose tetrad vector $n^a$ is defined from the 
dyad spinor $\iota^A$ by $n^a := \iota^A \ovl \iota^{A'}$.  This subset of the 
future null cone can be visualized as its intersection with a null hyperplane 
defined by $t + z = \sqrt{2}$ in the local Minkowski coordinates.  To the mind 
accustomed to Euclidean geometry, this intersection might be assumed  
to be a paraboloid.  However, interestingly, the Lorentzian structure 
of the spacetime metric causes the intersection to be, in terms of the 
induced metric, a flat 
two-dimensional plane, with $\zeta$ a standard complex coordinate on this 
plane.  In fact, the absolute value of the degeneracy measure 
$\brak{\alpha_1}{\alpha_2}$, under this particular normalization condition, is 
precisely the quantity $\Delta_{12} = | \zeta_1 - \zeta_2 |$.  For this 
reason, the degeneracy measure used in~\cite{Campanelli2008} can be 
understood as a geometric distance between the two associated principal 
null directions along the cut made by a null hyperplane through the future 
null cone.  

The degeneracy measure $\Theta_{ij}$ introduced in Eq.~\eqref{e:Thetadefined} 
can similarly be understood in terms of the symplectic product 
$\brak{\alpha_1}{\alpha_2}$.  If the condition on the null vectors associated 
with the $\alpha_i^A$ is taken to be that $\vec V_i \cdot \vec e_0 = -1$, for 
a timelike tetrad vector defined from a Newman-Penrose tetrad through 
Eqs.~\eqref{e:orthonormal}, rather than $\vec V_i \cdot \vec n = -1$, then 
the spinors must be scaled as:
\beq
\alpha_i^A = \frac{1}{\sqrt{1 + \zeta_i \ovl \zeta_i}} \left( o^A + \zeta_i \iota^A \right).
\eeq
In this case the absolute value of $\brak{\alpha_1}{\alpha_2}$ becomes:
\beq
\left| \brak{\alpha_1}{\alpha_2} \right| = \frac{|\zeta_1 - \zeta_2|}{\sqrt{\left(1+\zeta_1 \ovl \zeta_1\right)\left(1+\zeta_2 \ovl \zeta_2\right)}},
\eeq
essentially equivalent to $\Theta_{12}$, as defined in 
Eq.~\eqref{e:Thetadefined}.

The distinction between the degeneracy measures $\Delta_{ij}$ and $\Theta_{ij}$ 
can therefore be understood as a distinction between two different realizations 
of the geometry of the space of null rays at a point.  If the geometry in this 
space is inferred by cutting the null cone with a null hyperplane, the distance 
function on the set of null rays is given by $\Delta_{ij}$.  If the cut is 
taken by a spacelike hyperplane, the distance is given by $\Theta_{ij}$.  

The ambiguity of these distance functions stems from the ambiguity of these 
cuts.  Fundamentally there are equally many degrees of ambiguity in both types 
of cut.  A spacelike hyperplane can be boosted in any of three directions, 
translating into three continuous degrees of ambiguity for $\Theta_{ij}$ at 
each point in spacetime.  A null hyperplane cut can also be given 
in terms of three degrees of freedom: the null normal to the hyperplane (for 
which there are two degrees of freedom, the anti-celestial sphere), and a 
parameter describing the translation of the hyperplane away from the 
vertex of the cone.  This last degree of freedom also exists for spacelike 
hyperplanes, but because the intersection of the spacelike hyperplane with the 
null cone is compact (specifically a two-sphere), one can fix this translation 
degree of freedom by fixing the area of the sphere.  In the case of a cut by 
a null hyperplane, the intersection is noncompact, so this degree of freedom 
cannot be fixed.

Though the degeneracy measure $\Theta_{ij}$ may be no more well-defined in 
general than $\Delta_{ij}$, there are still reasons to prefer it for purposes 
of defining a notion of approximate Petrov class.  The main reason is that 
when a null hyperplane cut is made through a null cone, one special null ray 
is singled out: the one parallel to the hyperplane.  Again, the 
intersection of the future null cone with a null hyperplane is itself a 
spacelike two-dimensional plane, and because the null ray parallel to the 
hyperplane never intersects the hyperplane, it is only represented on 
the intersection plane as a point at infinity.  Eq.~\eqref{e:nullcut} shows 
that the null ray that gets mapped to the point at infinity is the one that 
points along the tetrad $\vec n$ vector.  This is the behavior 
that we saw in Sec.~\ref{s:quasikinnersley}.  The quasi-Kinnersley tetrad 
naturally adapts itself to the principal null directions in the ringdown 
to Kerr geometry, such that two of them fall toward the origin of the 
$\zeta$ plane and two approach infinity.  This is because the 
quasi-Kinnersley tetrad 
is {\em designed} to adapt itself to nearly-degenerate principal null 
directions.  To the extent that the numerical tetrad approximates a 
quasi-Kinnersley tetrad (a common implicit hope for the extraction of 
gravitational waveforms) this behavior will be seen also in numerical 
simulations.  An example of this will be seen in the next section.

\section{Numerical Results}
\label{s:results}

Our numerical implementation of these mathematical tools 
begins, as in Ref.~\cite{Campanelli2008}, with the fourth-order polynomial 
in Eq.~\eqref{e:polynomial}.  We begin by computing the Weyl scalars in a 
reference tetrad.  The timelike orthonormal tetrad leg $\vec e_0$ 
is taken to be the normal to the spatial slice, and the spacelike orthonormal 
tetrad legs are constructed from a Gram-Schmidt orthogonalization of the basis 
vectors of a spherical-like coordinate system within the spatial slice, 
essentially similar to the method in~\cite{Campanelli2008}.  This tetrad is 
singular at the $z$ axis, as the complex phase of the $\vec m$ leg becomes 
undefined due to the coordinate singularity of the spherical chart, but all 
quantities we present will be independent of this complex phase, and thus will 
have well-defined values on the axis.  

Our 
code computes the electric and magnetic parts of the Weyl curvature tensor 
directly from data on the spatial slice, using Gauss-Codazzi relations and 
assuming the Einstein equations are satisfied and that no matter fields are 
present:
\beqa
E_{ij} &=& \left({}^3R_{ij} + K K_{ij} - K_{ik} K_j^k \right)_{\rm STF}\\
B_{ij} &=& \left(\epsilon_i{}^{mn}D_mK_{nj}\right)_{\rm STF}.
\eeqa
Here, $K_{ij}$ is the extrinsic curvature of the spatial slice, $D_i$ is the 
torsion free covariant derivative compatible with the spatial metric, 
${}^3R_{ij}$ its Ricci curvature, $\epsilon_{ijk}$ the spatial Levi-Civita 
tensor, and the subscript ${\rm STF}$ means that the quantity in brackets is 
made symmetric and tracefree in the indices $i$ and $j$.  Once these tensors 
are computed, we construct the Weyl scalars from them as in Eqs.~(4) 
of Ref.~\cite{Nerozzi2007}, using the radial tetrad leg for $\vec u$, and the 
polar leg plus $i$ times the azimuthal leg for $\vec m$.  

Once the Weyl scalars are known, one can go about solving for the roots 
$\lambda_i$ of Eq.~\eqref{e:polynomial}.  We do so point by point on the 
computational grid with simple Newton-Raphson iteration and polynomial 
deflation~\cite{numrec_cpp}.  In many cases, these methods are conventionally 
followed by root polishing --- using the computed roots of the deflated 
polynomials as initial guesses in new Newton-Raphson iterations of the initial 
polynomial, with the hope of correcting roundoff error accumulated in the 
deflation process --- but in this case root polishing has no noticeable 
effect.  This is presumably because the roots under consideration are very 
nearly 
degenerate, so error in the Newton-Raphson iterations themselves dominates the 
error accumulated in the polynomial deflation.  

As in Ref.~\cite{Campanelli2008}, we focus our attention on a simulation 
of the 
ringdown of a binary black hole merger to Kerr geometry.  The simplest example 
of such a merger is one following the inspiral of two equal mass nonspinning 
black holes in a noneccentric configuration.  This data set was presented 
in detail in Ref.~\cite{Scheel2008}, and the multipolar structure of the 
post-merger horizon was studied in Ref.~\cite{Owen2009}.  In the 
former paper, it was noted that two independent measures of black hole spin, 
designed to agree if the final black hole is Kerr, agreed to well within the 
estimated accuracy of the numerical truncation.  In the latter paper, this 
correspondence 
was studied in much greater detail, demonstrating that all of the multipole 
moments on the apparent horizon that we were able to compute agreed very 
well with those of the Kerr horizon (see 
Refs.~\cite{Schnetter2006, Jasiulek2009} for a similar analyses).  While these 
provide a very compelling 
case that the final black hole is Kerr, they do not present a major benefit of 
the methods described here and in Ref.~\cite{Campanelli2008}: being fully 
local.  The degeneracy measures described here can be 
computed independently at each point in spacetime, rather than simply for the 
apparent horizon as a whole.  In 
this way one can imagine demonstrating not just the fact that a spacetime is 
settling down to Kerr geometry, but where it is doing so more quickly and more 
slowly, and possibly even the relationship between the approach to Kerr 
geometry and the presence of gravitational radiation.  

This locality of the approximate Petrov classification system, while 
beneficial for the reasons described above, unfortunately comes at the cost 
of another type of gauge ambiguity.  If one wishes to investigate the 
time dependence of the degeneracy measures, then one 
must choose a worldline in spacetime along which to compute these quantities.  
In principle one could reduce this ambiguity, for example by computing along 
timelike geodesics, or worldlines preferred by some sort of symmetry, if any 
exist.  For example, the symmetries inherent in a merger of equal mass, 
initially nonspinning holes 
provide at least one preferred axis for consideration.  The initial data 
satisfy a discrete symmetry under 180-degree rotations about a certain axis, 
taken in our simulations to be the coordinate $z$ axis, along which the 
initial ``orbital angular 
momentum vector'' can be intuitively said to point.  To the extent that the 
numerical simulation preserves this discrete symmetry, the $z$ axis sweeps out 
a geometrically well-defined worldsheet in spacetime.  In principle this 
timelike worldsheet could be restricted to a single well-defined timelike 
worldline, on which data can be extracted, by intersecting the worldsheet with 
a level surface of some curvature invariant.  Here, however, we do not go to 
such lengths, electing instead to follow coordinate worldlines, as 
in Ref.~\cite{Campanelli2008}, but paying special attention to the symmetry 
axis.  

\begin{figure}
\begin{center}
\includegraphics[scale=.3]{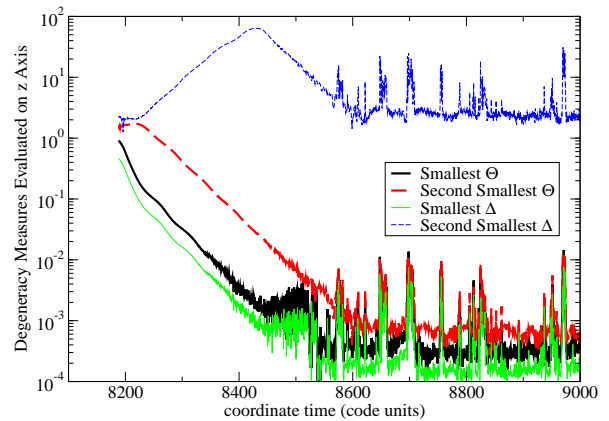}
\end{center}
\caption{ \label{f:zaxis} The ringdown after the merger of two equal mass, 
initially nonspinning holes.  The curves show the behavior of the two smallest 
values of each of the degeneracy measures $\Delta_{ij}$ and $\Theta_{ij}$, 
with respect to coordinate time, evaluated at $z = 4.5$, $x = y = 0$ in an 
asymptotically inertial coordinate system.  The heavier curves show the two 
values of the $\Theta_{ij}$ measure, the lighter curves the $\Delta_{ij}$ 
measure.  The solid curves show the smaller values of these measures, and 
the dashed curves the larger.  The symmetries along this axis 
force the tetrad to satisfy the basic conditions of a ``quasi-Kinnersley'' 
tetrad, as described in section~\ref{s:quasikinnersley}, and the results of 
that section explain the initial exponential growth of the higher 
$\Delta_{ij}$ curve.  
}
\end{figure}

In Fig.~\ref{f:zaxis}, 
data are shown for the two smallest --- smallest 
among the various possible root pairings --- values of the degeneracy measures 
$\Delta_{ij}$ and $\Theta_{ij}$ evaluated at the coordinate location 
$x = y = 0$, $z = 4.5$\footnote{For a sense of scale we note that the 
apparent horizon, in these coordinates, settles down at late times roughly to 
a coordinate sphere with radius of $2.61$.} 
as a function of coordinate time after the formation of the common 
apparent horizon in the dataset described in Ref.~\cite{Scheel2008}.  Because 
the tetrad is, by construction, adapted 
to the symmetry axis at this location,\footnote{Actually the tetrad is, 
strictly speaking, not well defined on the axis, because it is constructed 
from the spherical coordinate basis, which is singular there.  However, for 
the objects we compute, this singularity has no effect.  The $\vec \ell$ and 
$\vec n$ tetrad legs are well-defined on the axis, it is just the complex 
phasing of the $\vec m$ vector that becomes undefined there.  The actual 
quantities we compute, however, $\Delta_{ij}$ and $\Theta_{ij}$, are invariant 
under spin transformations (spin-boost transformations with $|c| = 1$ in 
Eq.~\eqref{e:spinboost}).  Because there are no grid points on the axis, these 
quantities can always be computed, and because they are spin invariant, they 
can be smoothly interpolated to the axis.} it is forced to be ``transverse'' 
in the sense of Sec.~\ref{s:quasikinnersley} (a fact which we have verified 
by a direct inspection of the computed values of $|\Psi_1|$ and $|\Psi_3|$).  
In Fig.~\ref{f:zaxis}, we initially see exponential growth in the 
second-smallest root difference $\Delta_{ij}$, as one would expect from the 
considerations of Sec.~\ref{s:quasikinnersley}.
Eventually, this exponential growth 
gives way to exponential decay, similar to the behavior seen in 
Fig.~\ref{f:qKoffset_manyangles}.  This occurs because the data we compute 
here are actually interpolated to the polar axis from data on grid points 
slightly offset from it.  On these grid points, the tetrad differs slightly 
from the quasi-Kinnersley tetrad, as in the discussion near the end of 
Sec.~\ref{s:quasikinnersley}.  One 
might hope that this eventual decay would only occur on these offset grid 
points, and that the data interpolated to the axis would grow indefinitely as 
$\sqrt{-6 \Psi_2/\Psi_4}$, but as the black hole settles down the 
growing peak in Fig.~\ref{f:qKoffset_manytimes} shrinks in width, so 
eventually one would expect it not to be resolved by the spectral 
discretization.  
Incidentally, we have confirmed that 
the rates of exponential decay in the decaying curves, and the rate of 
exponential growth in the growing curve, each roughly equal half of the 
damping rate of the least-damped quasinormal mode of a Kerr black hole of the 
same final mass and spin as our final remnant.  One would expect this 
from Eq.~\eqref{e:halfrate}.  The most important issue to note about 
Fig.~\ref{f:zaxis}, though, is the discrepancy between the picture implied by 
the $\Delta_{ij}$ values, and that implied by the $\Theta_{ij}$ values.  The 
highest and lowest curves are the two relevant values of $\Delta_{ij}$.  At 
early times, even the qualitative behavior of these curves are different, 
one growing and one decaying.  Even at late times, when both curves decay 
exponentially (and eventually settle to fixed limits due to numerical 
truncation error), they still differ by four orders of magnitude.  If 
$\Delta_{ij}$ were naively interpreted as defining the ``nearness'' to any 
Petrov class, then the response to Fig.~\ref{f:zaxis} would be that the 
final result of the numerical simulation is of Petrov Type II, not Type D, on 
this axis.  The other two curves in Fig.~\ref{f:zaxis} tell a very different 
story.  The two heavier curves in Fig.~\ref{f:zaxis} show the 
two smallest values of the $\Theta_{ij}$ measure.  Both curves fall 
exponentially at the same rate, and they lie within roughly a factor of ten 
of one another throughout the entire ringdown.  According to $\Theta_{ij}$ the 
spacetime falls quite unambiguously to Petrov Type D on the polar axis.  

\begin{figure}
\begin{center}
\includegraphics[scale=.3]{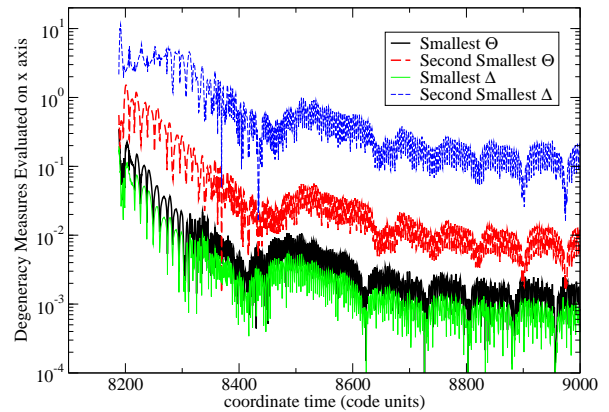}
\end{center}
\caption{ \label{f:xaxis} The two smallest values of the degeneracy measures 
$\Delta_{ij}$ and $\Theta_{ij}$ evaluated at the coordinate location 
$x = 4.5$, $y = z = 0$ in the ringdown after a merger of equal-mass, 
initially nonspinning black holes, as in Fig.~\ref{f:zaxis}.  The lighter 
curves, which are generally the highest and lowest curves, are the two values 
of the $\Delta_{ij}$ measure.  The heavier curves represent the $\Theta_{ij}$ 
measure.  Initial growth 
of the larger $\Delta_{ij}$ value is still present, but less striking than in 
Fig.~\ref{f:zaxis}.  Nonetheless, the two $\Delta_{ij}$ values again 
differ by multiple orders of magnitude, while the two $\Theta_{ij}$ values 
generally differ by only one.
}
\end{figure}

\begin{figure}
\begin{center}
\includegraphics[scale=.3]{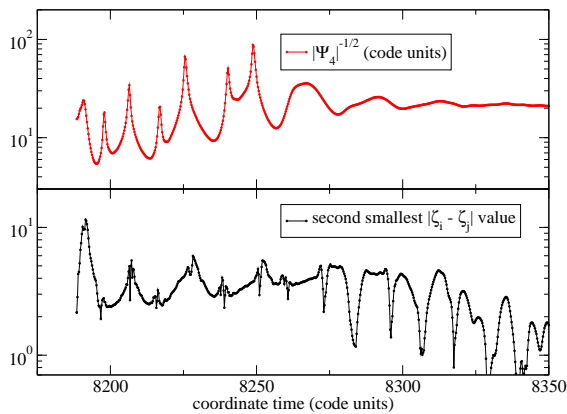}
\end{center}
\caption{ \label{f:xaxis_Diff2zoom} Magnification of the highest curve in 
Fig.~\ref{f:zaxis}.  The sharp peaks in the bottom curve before time $8275$ 
imply that principal 
null directions are occasionally crossing the tetrad legs.  The upper curve 
corroborates this by explicitly showing that the absolute value of the 
Weyl scalar $\Psi_4$ approaches zero at times coincident with these sharp 
peaks.
}
\end{figure}

The behavior at different coordinate locations is less striking, but shows 
roughly similar features.  Fig.~\ref{f:xaxis} 
presents the same quantities as 
Fig.~\ref{f:zaxis} computed instead at $x = 4.5$, $y = z = 0$.  Again, 
the highest and lowest curves are the two relevant 
values of $\Delta_{ij}$.  The higher one grows slightly (on average) in the 
early ringdown, but settles into exponential decay quite a bit sooner than in 
Fig.~\ref{f:zaxis}, and throughout the ringdown remains separated from the 
values in the lowest curve by roughly two to three orders of magnitude.  This 
still, however, provides a marked contrast from the two $\Theta_{ij}$ curves, 
which again lie within roughly a single order of magnitude of one another 
throughout the ringdown.  A particularly interesting feature is visible in the 
uppermost curve of Fig.~\ref{f:xaxis} from the beginning of the post-merger 
dataset to roughly the coordinate time $8275$.  This time frame is magnified 
in Fig.~\ref{f:xaxis_Diff2zoom}.
Because the spacetime is symmetric under reflections across the $z = 0$ plane, 
the spatial projections of the principal null directions on this plane must 
either be tangent to the plane, or otherwise reflection-symmetric across it.  
The jagged peaks in Fig.~\ref{f:xaxis_Diff2zoom} imply that the former 
possibility seems to apply here.  When the tetrad $\vec n$ vector happens 
to point along a principal null direction, the corresponding $\zeta$ value 
of the polynomial root becomes infinite.  If the spatial projections of two 
of the principal null directions lie in the same plane as the spatial 
projection of the tetrad $\vec n$ vector, and they oscillate in direction, 
repeatedly crossing the spatial projection of $\vec n$ (due either to 
physical or gauge effects), then one would expect the $\Delta_{ij}$ values 
involving these principal null directions to show sharp, repeating peaks, such 
as those in Fig.~\ref{f:xaxis_Diff2zoom}.  Eventually, such crossings could be 
expected to stop as the angle between the spatial projections of the principal 
null directions becomes small and as gauge and physical oscillations become 
less wild.  
After roughly $t = 8275$ in these code units\footnote{To aid in translating 
the code units to physically relevant units, we note that the final mass of 
remnant black hole, in these code units, is roughly $1.98$.} the 
oscillations in this curve become more smooth, implying that the principal 
null directions are no longer crossing $\vec n$.

\begin{figure}
\begin{center}
\includegraphics[scale=.3]{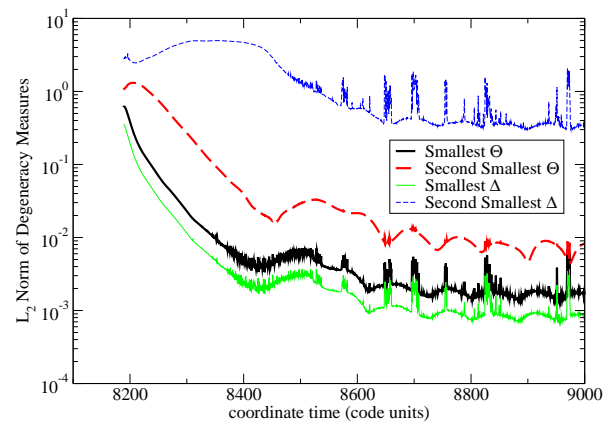}
\end{center}
\caption{ \label{f:L2Norms} $\L_2$ norms of the two smallest values of 
$\Delta_{ij}$ and $\Theta_{ij}$, integrated over a thick spherical shell 
extending from just within the apparent horizon, $r = 2.2$ in code 
coordinates, to an outer boundary at $r = 5$ code coordinates.  Again, 
both values of $\Theta_{ij}$ 
fall quickly to zero at the same rate, while the larger $\Delta_{ij}$ value 
hangs up initially, and eventually falls only to a level over two orders of 
magnitude above its smaller counterpart.  
}
\end{figure}

Figure~\ref{f:L2Norms} 
presents $L_2$ norms of the same degeneracy measures for the same ringdown 
dataset.  
While this avoids the choice of a specific coordinate location for 
observation, we should note that there is still a certain amount of coordinate 
ambiguity in this quantity.  The $L_2$ norm is only over a certain subset of 
the 
spatial slice.  The inner boundary is the excision boundary of the simulation, 
slightly inside the apparent horizon.  The outer boundary is a boundary 
of subdomains in the simulation, with coordinate radius 5.  The 
purpose of this outer boundary is to 
avoid numerical difficulties when the Weyl scalars become too small to 
calculate accurately the roots of the polynomial in Eq.~\eqref{e:polynomial}.  
Again in Fig.~\ref{f:L2Norms}, we see the larger value of $\Delta_{ij}$ 
hanging up while both values of $\Theta_{ij}$ decay exponentially.

\section{Discussion}
\label{s:discussion}

The primary goal of this paper has been to explain the peculiar behavior, noted 
in 
Ref.~\cite{Campanelli2008}, that the spacetimes of binary black hole mergers 
seem to ``hang up'' in Petrov Type II, and if they fall to Type D at all 
(according to one's choice of tolerance), they do so much later.  

Properly clarifying this issue has required us to investigate the 
geometrical meaning of the polynomial in Eq.~\eqref{e:polynomial} and the 
measure of degeneracy principally used in Ref.~\cite{Campanelli2008}, the 
absolute value of the difference of any two roots, 
$\Delta_{ij} := |\lambda_i - \lambda_j|$.  The true space of interest is the 
space of future-directed null rays at a point, the so-called anti-celestial 
sphere.  As argued in Sec.~\ref{s:spinors}, the complex quantity $\lambda$ 
acts as a coordinate on this two-dimensional space.  It should therefore not 
be surprising that $\Delta_{ij}$, a coordinate distance, fails to represent 
geometries in the space of null rays, since it is impossible to cover a 
topological sphere with a single coordinate chart without the presence of 
coordinate singularities.  

It would therefore seem that the right thing to do would be to consider 
truly geometrical distances in the space of null rays as defining the nearness 
of principal null directions to one another.  This approach, unfortunately, is 
clouded by the nonexistence of a preferred metric structure on the 
anti-celestial sphere.  The anti-celestial sphere has a six-dimensional 
conformal group, corresponding to the proper Lorentz group of 
Minkowski spacetime.  While this group carries a three-dimensional subgroup of 
isometries --- corresponding to rotations --- which have no effect on 
the ``distances'' between any two null rays, the three remaining 
dimensions --- corresponding to boosts --- conformally 
rescale the metric on the space of null rays.  
For this reason, fixing a unique geometry on the space of null rays requires 
fixing a unique ``observer'' with respect to which this boost freedom is 
fixed.  In the introduction, we described similar difficulties in attempting 
to define a concept of approximate Petrov class by algebraic means.  

In Sec.~\ref{s:spinors} we also presented a geometrical interpretation of the 
quantity $\Delta_{ij}$.  Rather than simply as a coordinate distance on the 
space of null rays, $\Delta_{ij}$ can be interpreted as a metric distance 
along the cut of a null cone 
made by a null hyperplane.  In a sense, one is here adapting the metric on 
the space of null rays to a null observer.  Similarly, $\Theta_{ij}$ is a 
distance function on the intersection of the future null cone with a spacelike 
plane orthogonal to our timelike observer.  

There are equally many degrees of freedom in cutting the null cone with a 
spacelike hyperplane as there are in cutting it with a null hyperplane (once 
one restricts the spacelike cuts to normalize the area of the anti-celestial 
sphere, a restriction that cannot be made on null hyperplanes because the 
intersection is noncompact).  For this reason the degeneracy measure that we 
have introduced, $\Theta_{ij}$ in Eq.~\eqref{e:Thetadefined}, cannot be 
considered fundamentally any more invariant than $\Delta_{ij}$, though in 
practice it is easier to imagine a fleet of preferred timelike observers than 
of null observers, such as stationary observers in stationary spacetimes, 
observers adapted to timelike approximate Killing vectors in approximately 
stationary spacetimes (if such vectors can be reasonably defined), 
or freely falling observers following timelike geodesics from infinity.  
We have not, however, attempted to find any such preferred classes of 
observers in the numerical results presented here, either for fixing the 
geometry on the space of null rays or for fixing the worldlines along which 
data are extracted.  The main value of the new degeneracy measure 
$\Theta_{ij}$ is not that it is more gauge-invariant, but rather that it 
naturally captures the compactness of the space of null rays, 
and thereby avoids relegating any particular null ray to a point at infinity.  

The need to avoid relegating any null ray to a point at infinity is 
particularly acute in practice, as the rays at infinity in 
the physically preferable quasi-Kinnersley tetrads become the principal null 
directions themselves as a spacetime settles down to Kerr geometry.  This 
behavior was investigated in Sec.~\ref{s:quasikinnersley}.  As principal null 
directions relax to degeneracy at the point at infinity in $\lambda$ space, 
the degeneracy measure $\Delta_{ij}$ grows exponentially rather than decaying 
exponentially as one would naively expect.  While the tetrads used in 
numerical simulations are not commonly quasi-Kinnersley tetrads, there is 
generally an implicit hope, for purposes of wave extraction, that they are 
close to it in some rough sense.  Indeed, as is visible in Fig.~\ref{f:zaxis}, 
this nonintuitive behavior 
in the quasi-Kinnersley tetrad can corrupt measurements of $\Delta_{ij}$ in 
even a simple coordinate-adapted tetrad ({\em cf.} 
Fig.~\ref{f:qKoffset_manyangles}).  

The other figures in Sec.~\ref{s:results} tell much the same story, though in 
somewhat less dramatic terms.  Fig.~\ref{f:xaxis_Diff2zoom} shows indications 
of principal null directions directly crossing the tetrad null vectors, 
repeatedly causing the null directions to be represented by the point at 
infinity in $\lambda$ space, due purely to the choice of spatial tetrad.  
Fig.~\ref{f:L2Norms} shows that the hangups in the degeneracy measure 
$\Delta_{ij}$ are not limited to the partially symmetry-preferred 
$x$ and $z$ axes.  In fact, $\Delta_{ij}$, computed everywhere in a tetrad 
adapted to spherical coordinates, clearly shows this hangup in Petrov Type II 
even in an integral $L_2$ norm, while $\Theta_{ij}$ does not.  

Another motivation of this paper has been to provide further evidence that the 
final remnants of our black hole merger simulations are Kerr black holes.  
This was indeed the central focus of Ref.~\cite{Campanelli2008}, and they 
even went so far as to demonstrate that their final black hole has no NUT 
charge or acceleration (as in the C-metrics; see 
Ref.~\cite{GriffithsPodolsky}), an issue that we have not explored here.  

The question of whether a black hole is ``settling down to Kerr'' can be 
attacked at a variety of levels.  In a recent paper~\cite{Owen2009}, we 
studied the question 
at a quasilocal level, verifying that the quasilocal source multipole moments 
of the black hole settle to the values required by the Kerr geometry (see also 
Refs.~\cite{Schnetter2006, Jasiulek2009}).  
More recently, Ref.~\cite{BaeckdahlKroon2010} presented theoretical tools 
for approaching the question at a global level (global on any given spacelike 
slice).  The approach taken in Ref.~\cite{Campanelli2008} was in part 
local (measurement of $\Delta_{ij}$), and in part global.  The method by 
which Campanelli {\em et al.} verified the vanishing of the NUT charge and 
acceleration involved limits of curvature invariants to large radii.  If one 
wishes to rule out NUT charge or acceleration at a local level, to provide a 
more cohesive picture when combined with local algebraic degeneracy measures, 
this can be done with quantities presented in Ref.~\cite{FerrandoSaez2009}, 
though as described in the introduction of this paper, collapsing 
tensorial quantities to scalar quantities would require 
a positive-definite background metric, which could require fixing a slicing 
or a threading of spacetime.

\acknowledgments{
I thank Mark Scheel for providing the numerical simulation data studied in 
Sec.~\ref{s:results}.  I also thank the Caltech and Cornell numerical 
relativity groups, particularly Jeandrew Brink, Larry Kidder, Geoffrey 
Lovelace, and Saul Teukolsky for frequent discussions.  This work is supported 
in part by grants from the Sherman Fairchild Foundation to Cornell, and by 
NSF grants PHY-0652952, DMS-0553677, PHY-0652929, and NASA grant NNX09AF96G.
}

\bibliography{References}

\end{document}